\documentclass[preprint2]{aastex}
\begin{document}
\bibliographystyle{unsrt}
%\draft
\title{Observational Constraints on the Role of the Crust
 in the Post-glitch Relaxation}
\author{M. Jahan-Miri}
\affil{Department of Physics, Shiraz University, Shiraz 71454,
Iran}
\authoremail{jahan@physics.susc.ac.ir}

\begin{abstract}
The observed large rates of spinning down after glitches in some
radio pulsars have been previously explained in terms of a
long-term spin-up behavior of a superfluid part of the crust of
neutron stars. We argue that the suggested mechanism is not
viable; being inconsistent with the basic requirements for a
superfluid spin-up, in addition to its quantitative disagreement
with the data. Hence, the observed post-glitch relaxations may not
be interpreted due only to the effects of the stellar crust.
\end{abstract}
\keywords{stars: neutron -- hydrodynamics -- pulsars}

\section{Introduction}
Glitches, and post-glitch relaxations, are widely believed to be
effects caused by (the superfluid component in) the crust of
neutron stars (see, eg., \markcite{LGS8} Lyne \& Graham-Smith
1998; \markcite{KLGJ03} Krawczyk~et. al. 2003). There exist,
however, observational data on glitches that could not be possibly
due to the role of the stellar crust. The data have been
previously explained  \markcite{APC} (Alpar, Pines \& Cheng 1990;
hereafter APC) in terms of a suggested spin-up of (a part of) the
crustal superfluid by the spinning down crust (``the container'')
over a time much larger than the associated relaxation timescale.
However, a closer look at the relative rotation of the superfluid
and its vortices reveals that the suggested mechanism fails
quantitatively by, at least, more than one order of magnitude. In
addition, the suggested spin-up process is also argued to be in
contradiction with the well-known requirements for a superfluid
spin-up. Hence, the (pinned) superfluid in the crust is not the
primary cause of the post-glitch relaxation. On the other hand,
the pinning of the superfluid vortices in the crust, and also in
the core, of a neutron star has recently been objected by some
authors on the account of the possible observations of long period
precession in isolated pulsars (Jones \& Anderson 2001; Link 2003;
Buckley, Metlitski \& Zhitnitsky 2004). Thus the post-glitch
relaxation must be driven by mechanism(s) other than that due only
to the crust superfluidity, whether the pinning is realized or
not. In section 2, the general role of an assumed superfluid
component in (the crust or in the core of) a neutron star on the
observable post-glitch behavior of the star is briefly described.
In section 3, the relevant observational data and the problem
raised by these observations against any model of post-glitch
relaxation based on the effects of the crust alone are
highlighted. The earlier suggested resolution \markcite{APC} (APC)
of the problem is then stated. In section 4, a quantitative
evaluation of the suggested mechanism is given, indicating a large
disagreement with the data. The subsection 4.1 presents a more
detailed discussion of the rotation of the different components in
the crust, paying particular attention to the vortex lines, and
arrives at the same conclusion as already deduced, in the section
4. In section 5, the feasibility of the suggested process, of
spinning up of a pinned superfluid component in the crust during
the spinning down of the crust itself, is questioned, altogether.
The possibility of such a process is argued to be ruled out, on a
general dynamical ground, and also according to the vortex creep
formulation. We conclude in section 6, with a speculative
suggestion for the possible cause of the observed effect.

\section{Crustal Superfluid: An Overview}
The spin-down rate $\dot\Omega_{\rm c}$ of the crust of a neutron
star, with a moment of inertia $I_{\rm c}$, obeys \markcite{BPP}
(Baym et~al. 1969b)
\begin{eqnarray}
 I_{\rm c} \dot\Omega_{\rm c} = N_{\rm em} -
                \Sigma I_{\rm i}\dot\Omega_{\rm i}
\end{eqnarray}
where $N_{\rm em}$ is the negative electromagnetic torque on the
star, and $\dot\Omega_{\rm i}$ and $I_{\rm i}$ denote the rate of
change of rotation frequency and the moment of inertia of each of
the separate components, respectively, which are summed over.
Steady state implies \(\dot\Omega_{\rm i} =\dot\Omega_{\rm c} =
\dot \Omega \equiv {N_{\rm em} \over I}\), for all $i$, where \(
I= I_{\rm c}+\Sigma I_{\rm i}\) is the total moment of inertia of
the star. Different models for the post-glitch recovery, and in
particular the model of vortex-creep, invoke a
decoupling-recoupling of a superfluid component in the {\em crust}
of neutron stars \markcite{AAP,J91a,EL}(Alpar et al. 1984; Jones
1991a; Epstein, Link \& Baym 1992). The rest of the star including
the core (superfluid) is assumed in these models to be
rotationally coupled to the non-superfluid constituents of the
crust, on timescales much shorter than that resolved in a glitch.

The role of any superfluid component of a neutron star in its
post-glitch behavior is understood as follows. Spinning down (up)
of a superfluid at a given rate is associated with a corresponding
rate of outward (inward) radial motion of its vortices. If
vortices are subject to pinning, as is assumed for the superfluid
in the crust of a neutron star, a spin-down (up) would require
unpinning of the vortices. This may be achieved under the
influence of a Magnus force $\vec{F}_{\rm M}$ acting on the
vortices, which is given, per unit length, as \markcite{S89}
\begin{eqnarray}
\vec{F}_{\rm M} & = & -  \rho_{\rm s} \vec{\kappa} \times
                                (\vec{v}_{\rm s} - \vec{v}_{\rm L})
\end{eqnarray}
where $\rho_{\rm s}$ is the superfluid density, $\vec{\kappa}$ is
the vorticity of the vortex line directed along the rotation axis
(its magnitude $\kappa = { h \over 2 m_{\rm n}}$ for the neutron
superfluid, where $m_{\rm n}$ is the mass of a neutron), and
$\vec{v_{\rm s}}$ and $\vec{v_{\rm L}}$ are the local superfluid
and vortex-line velocities. Thus, if a lag \( \omega \equiv
\Omega_{\rm s} - \Omega_{\rm c}\) exist between the rotation
frequency $\Omega_{\rm s}$ of the superfluid and that of the
vortices (pinned and co-rotating with the crust) a radially
directed Magnus force \((F_{\rm M})_r= \rho_{\rm s} \kappa r
\omega \) would act on the vortices, where $r$ is the distance
from the rotation axis, and \(\omega >0 \) corresponds to an
outward directed $(F_{\rm M})_r$, vice-versa. The crust-superfluid
may therefore follow the steady-state spin-down of the star by
maintaining a critical lag $\omega_{\rm crit}$ which will enables
the vortices to overcome the pinning barriers. The critical lag is
accordingly defined through the balancing $(F_{\rm M})_r$ with the
pinning forces.

At a glitch a sudden increase in $\Omega_{\rm c}$ would result in
\( \omega<\omega_{\rm crit}\), hence the superfluid becomes
decoupled and could no longer follow the spinning down of the star
(ie. its container). If, as is suggested \markcite{AI}, the glitch
is due to a sudden outward release of some of the pinned vortices
the associated decrease in $\Omega_{\rm s}$ would also add to the
decrease in $\omega$, in the same regions. Therefore a fractional
increase ${\Delta \dot \Omega_{\rm c} \over \dot \Omega_{\rm c}}$
same as the fractional moment of inertia of the decoupled
superfluid would be expected. This situation will however persist
only till \( \omega=\omega_{\rm crit}\) is restored again (due to
the spinning down of the crust) and the superfluid re-couples, as
illustrated by $\Omega_{\rm s}$ and $\Omega_{\rm c}$ curves in
Fig.~1. The vortex creep model suggests \markcite{AAP} (Alpar
et~al 1984) further that a superfluid spin-down may be achieved
even while \( \omega < \omega_{\rm crit}\), due to the creeping of
the vortices via their thermally activated and/or quantum
tunnelling movements. A superfluid spin-down with a steady-state
value of $\omega < \omega_{\rm crit}$, and also a post-glitch
smooth gradual turn over to the complete re-coupling for each
superfluid layer is thus predicted in this model. It may be
however noticed that the predicted gradual (exponential) recovery
of the crustal spin frequency is not an effect peculiar to the
creep process. It is mainly caused due to the assumed varying
amplitude of the glitch-induced jump in $\Omega_{\rm s}$ in the
different layers of the superfluid (corresponding to their assumed
varying critical lag values), which are thus re-coupled at various
times. A similar behavior should be also expected even in the
absence of any creeping, given the same series of the superfluid
layers. The induced $\Delta\dot\Omega_{\rm c} \over
\dot\Omega_{\rm c}$ during a superfluid decoupling phase according
to the creep model would be indeed the same as (or slightly
smaller than) otherwise.

\section{Observational Constraint}
The more recent glitches observed in Vela, and one in PSR~0355+54,
have shown values of
\begin{eqnarray}
{\Delta \dot \Omega_{\rm c} \over \dot \Omega_{\rm c}} > 10\%
\end{eqnarray}
with recovery timescales $\sim 0.4$~d, and $\sim 44$~d,
respectively \markcite{L87,APC,F95} (Lyne 1987; APC; Flanagan
1995). The data hence imply (Eq.~1) that a part of the star with a
fractional moment of inertia $> 10\%$ (up to $60\%$) is decoupled
from the crust during the observed post-glitch response. This is,
however, in sharp contradiction with the above glitch models,
since for the moment of inertia $I_{\rm crust}$ of the crustal
superfluid
\begin{eqnarray}
 {I_{\rm crust} \over I} \lesssim 2.5\%
\end{eqnarray}
\markcite{APC}(APC). The disagreement with the data
is indeed a fundamental shortcoming for the crustal models, and
not just a quantitative mismatch. Because, the predicted increase
for ${\Delta \dot \Omega_{\rm c} \over \dot \Omega_{\rm c}}$ in
these models is naturally bound to be smaller than the fractional
moment of inertia of the decoupled superfluid (also see the best
fit results of \markcite{AC3}Alpar et~al. 1993); the other
possibility raised in APC, to account for the larger spin-down
rates, is the point of issue in the following.

It has been suggested \markcite{APC}(APC) that the observed large
values of ${{\Delta \dot \Omega_{\rm c} \over \dot \Omega_{\rm
c}}} \sim 20 \%$ over a time scale $\tau_{\rm sp} \sim 0.4$~d, in
Vela, could be accounted for by assuming that part of the crustal
superfluid {\it spins up}, over the {\it same} time period
$\tau_{\rm sp}$ (see Fig.~1a). The superfluid would thus be
expected to act as a source of an additional spin-down torque on
the rest of the star and could, in principle, result in ${\Delta
\dot \Omega_{\rm c} \over \dot \Omega_{\rm c}}$ values much larger
than the fractional moment of inertia of the decoupled component.
For this to be realized, a region of the crust-superfluid with a
moment of inertia $I_{\rm sp}$ and a spin frequency $\Omega_{\rm
sp}$ has been assumed to support a tiny (positive) steady-state
lag $\omega_{\rm sp} \equiv \Omega_{\rm sp} -\Omega_{\rm c} \sim
3.5 \times 10^{-6} \ {\rm rad \ s}^{-1}$, in contrast to the much
larger steady-state value of the lag \( \omega \geq 10^{-2} \ {\rm
rad \ s}^{-1} \) in the rest of the crust-superfluid. Hence, a
glitch of a size \(\Delta\Omega_{\rm c} \sim 10^{-4} \ {\rm rad \
s}^{-1}\) would result in a ``reversed'' situation with \(
\Omega_{\rm c} >> \Omega_{\rm sp}\), which is further suggested
\markcite{APC} (APC) to be followed by a spinning up of the
superfluid over the time period $\tau_{\rm sp}$, as indicated in
Fig.~1a.

\section{A Quantitative Check}
The above spin-up scenario is however unable to account for the
observed effect, quantitatively. It assumes \markcite{APC} (APC)
that the total frequency difference ($\Omega_{\rm c}-\Omega_{\rm
sp}$), initially induced by the glitch, is slowly relaxed during
the period $\tau_{\rm sp}$. In contrast, we argue that only a
small fraction of the initial jump in the superfluid rotation rate
might be at all preserved for any such ``long-term'' spin-up
process. This is because the superfluid would be otherwise
rotating much slower than its vortices which are, by virtue of
their assumed pinning, co-rotating with the crust (see Fig.~1a).
That is, the rotational lag between the superfluid and its
vortices would be {\em much larger} than the associated critical
lag. If so, the pinning could not impede the vortex motions (see
above) and a fast superfluid spin-up should take place. It may be
recalled that the critical lag is, by definition, the minimum lag
required for the Magnus force on vortices to overcome the pinning
forces. When the instantaneous lag exceeds its critical value, the
pinning forces (in the azimuthal direction) would act as a major
source for the torque on the superfluid, resulting in a relaxation
even faster than in the absence of any pinning
\markcite{TT,ACG}(Tsakadze \& Tsakadze 1980; Adams, Cieplak \&
Glaberson 1985). In order to allow for the suggested {\it large}
frequency difference between the superfluid and the crust, while
maintaining a rotational lag (between the superfluid and its
vortices) {\em smaller} than the associated critical value, one is
forced to assume that the pinning is ``switched off'', which would
be in contradiction with the assumed pinning conditions.
Nevertheless, the superfluid relaxation for such free (unpinned)
vortices should still take place very quickly, as is further
discussed below.

Thus, the upper limit on the frequency difference $\Delta\Omega$
between the crust and the superfluid ($\Omega_{\rm sp}$), at the
beginning of the time period of interest $\tau_{\rm sp}$ and {\em
after} the ``fast'' early relaxation of the superfluid (discussed
above), would be $\Delta\Omega = \omega_{\rm sp} \sim
3.5\times10^{-6}\ {\rm rad~s}^{-1}$; see the discussion in \S4.1
below for a more detailed reasoning. This is the maximum
permissible frequency difference that one might, in principle,
consider to be further equilibrated between the crust and the
superfluid. In contrast, a value of $\Delta\Omega =
\Delta\Omega_{\rm c}\sim 1.3\times10^{-4}\ {\rm rad~s}^{-1}$ has
been adopted in APC. The corresponding time scale $\tau_{\rm sp}$,
for an assumed spin-up of the superfluid, may be estimated from
\markcite{BPP}(Basym et~al. 1969) (see also Eq.~2b in APC)
\begin{eqnarray}
       \left(\Delta\dot\Omega_{\rm c}\right)_{\rm sp}=
                {I_{\rm sp}\over I} { \Delta\Omega \over \tau_{\rm sp}}
\end{eqnarray}
where $(\Delta\dot\Omega_{\rm c})_{\rm sp}$ is the magnitude of
the change in $\dot\Omega_{\rm c}$ due to the spinning up of the
superfluid. Adopting the same parameter values as in APC, ie.
\({{(\Delta\dot\Omega_{\rm c})_{\rm sp}}/{\dot\Omega_{\rm c}}}=
0.2\), \(\frac{I_{\rm sp}}{I}=5.3\times10^{-3}\), \(\Delta\Omega=
\omega_{\rm sp}= 3.5\times10^{-6}\ {\rm rad~s}^{-1}\), and
$\dot\Omega_{\rm c}= 9.5\times10^{-11}\ {\rm rad~s}^{-2}$ for
Vela, Eq.~5 then sets an upper limit of
\begin{eqnarray}
\tau_{\rm sp}< 0.3 \ {\rm hr},
\end{eqnarray}
which is {\it too short} in comparison with the observed
timescales $\sim 0.4$~d. Hence, the crust-superfluid cannot be the
cause for the observed large spin-down rates even in the case of
1988 glitch of the Vela pulsar, addressed in APC. The disagreement
between the predicted and observed timescales would be even worse
for the case of 1991 glitch of the same pulsar, having observed
values of \({\Delta\dot\Omega_{\rm c}/{\dot\Omega_{\rm c}}} \sim
60\% \) over a similar relaxation time \markcite{F95}(Flanagan
1995). Also, an attempt to apply the same crust-superfluid spin-up
scenario to the case of PSR~0355+54 would result in more than {\it
three} orders of magnitudes difference between the predicted
$\tau_{\rm sp}$ and the observed relaxation time $\sim 40$~days.

\subsection{Further reasoning for \(\omega_{\rm sp}\), against
                  \(\Delta\Omega_{\rm c}\)}
The above, rather obvious, conclusion (about the proper value of
the frequency difference between the crust and the superfluid at
the beginning of the long term relaxation) may be further
explained by focussing attention on the behavior of the rotation
frequency $\Omega_{\rm L}$ of the {\it vortices}, in the region of
interest identified by $\Omega_{\rm sp}$. Since this has not been
explicitly specified in APC we, therefore, discuss the two
exclusive possibilities that might, in principle, arise and which
could be physically justified. Both cases lead, however, to the
same conclusion; intermediate cases for which no justification
exist should naturally fall in between (and there is no indication
in APC for any special effect due to such cases). It may be noted
that the cases to be considered should not be, however, paralleled
to the classification of (strong, weak, super-weak) pinning
regions, invoked in the literature on the vortex creep model. The
latter is based on the relative magnitude of the critical lag and
reflects the long term behavior of the superfluid relaxation
toward its steady state lag value. In contrast, the following two
cases concern the instantaneous response of the originally pinned
vortices upon a sudden jump in the rotation rate of the container
(see Fig.~1). The vortices might be spun up along with the crust
(container), and remain pinned {\em during} the sudden spin-up of
the container (case {\bf a}). Else, they could relax to a state of
co-rotation with the local superfluid, assuming the pinning to be
temporarily broken ({\bf b}). Hence, the rotational frequency of
the vortices in the region of interest would be such that either
\begin{eqnarray}
\Omega_{\rm L} &=& \Omega_{\rm c} \ \ \ \ \ {\rm (pinning \ conditions), or \  else} \\
\Omega_{\rm L} &=& \Omega_{\rm sp} \ \ \ \ \ {\rm (Helmholtz \
theorem)},
\end{eqnarray}
just {\it at} the beginning of the interval $\tau_{\rm sp}$,
namely after the jump in $\Omega_{\rm c}$ (the observable glitch)
has been accomplished (see Fig.~1a),

In the case {\bf (a)}, the superfluid ($\Omega_{\rm sp}$) must
have also been spun up, along with the crust and the vortices, to
(at least) a state such that \( \Omega_{\rm sp}-\Omega_{\rm c}=
-\omega_{\rm sp}\) (contrast with the scenario adopted by APC as
depicted in Fig.~1a). Otherwise the pinning would be broken
(contrary to the assumption) by the associated radial Magnus force
on the vortices. That is, if the superfluid rotation rate is
assumed to retain its pre-glitch value, while the pinned vortices
have been spun up along with the crust, the instantaneous
(negative) lag between the superfluid and its vortices far exceeds
its critical value. Under such conditions, the superfluid
(spin-up) relaxation could not be impeded by the pinning forces.
The relaxation would occur on short timescales similar to the case
of free vortices, also according to the vortex creep model (see,
eg., the discussion related to Eq.~11 in Alpar et~al. 1984). Such
a fast relaxation of the superfluid in the crust of a neutron
star, when the lag exceeds its critical value, is in fact invoked
in the vortex creep model as the cause of the glitches
\markcite{AI,AAP}(Anderson \& Itoh 1975; Alpar et~al. 1984). Given
the existing (two minute) upper limit on the rise time of the
glitch \markcite{F95}(Flanagan 1995), the superfluid spin-up (for
$\Omega_{\rm sp}$, in the present case, until its rotational lag
with its vortices drops to values smaller than the critical value)
has to be likewise accomplished on a time scale of a minute
(contrast with $\tau_{\rm sp}$). Formally, this may be verified
from the following relation which is prescribed, in the creep
model, for such cases when the value of the lag $\omega
>> \omega_{\rm crit}$ \markcite{CAP}
\begin{eqnarray}
v_r = v_0 \exp{[- \frac{E_{\rm p}}{{\rm k}T} \left(
\frac{\omega_{\rm crit}-\omega}{\omega_{\rm crit}}\right)] /
\left\{ \exp[- \frac{E_{\rm p}}{{\rm k}T} \left(\frac{\omega_{\rm
crit}-\omega}{\omega_{\rm crit}}\right)]+1\right\}},
\end{eqnarray}
which amounts to
\begin{eqnarray}
v_r \sim v_0,
\end{eqnarray}
where $v_r$ is the vortex radial velocity, and $v_0\sim10^7 \ {\rm
cm}/{\rm s}$  corresponds to a spin-up (down) time scale $\sim
0.1$~s. Indeed, laboratory experiments on superfluid Helium have
also showed \markcite{T75,A87,TT} (Tsakadze \& Tsakadze 1975;
Alpar 1987; Tsakadze \& Tsakadze 1980) that a pinned superfluid
either spins up {\em along} with its vessel, or it never does so
during the subsequent spinning down of the vessel (for conditions
corresponding to $|\Omega_{\rm sp} -\Omega_{\rm L}| < \omega_{\rm
sp}$).

The case {\bf (b)}, on the other hand, is not in accord with the
general pinning conditions assumed in the vortex creep model, and
is not likely to be invoked in that context. Nevertheless, the
superfluid spin-up in the crust of a neutron star, for such a case
of free unpinned vortices, is again expected to occur over very
short timescales. The longest timescale for the spin-up of the
crust by freely moving vortices, due to nuclear scattering alone,
has been estimated \markcite{EL}(Epstein et~al. 1992) to be only
$\lesssim 5$~s, for the Vela pulsar. It is only natural that the
conclusions of the above two cases are the same, as the pinned
vortices should behave like the free ones, once the critical lag
is exceeded.

Therefore, and in either cases ({\bf a} or {\bf b}), the
superfluid would be spun up within a period of {\em only few
seconds} to (at least) a frequency such that \(\Omega_{\rm
sp}-\Omega_{\rm c}=-\omega_{\rm sp}\), while the observable jump
in $\Omega_{\rm c}$ takes place at the glitch (see Fig.~1b). It is
noted that the steady-state lag $\omega_{\rm sp}$ according to the
vortex creep model, in the so-called non-linear regime, would be
slightly smaller than the critical lag (though slightly larger in
the absence of any creeping). The difference is however a tiny
fraction of the critical lag \markcite{AAP}(Alpar et~al. 1984),
and may therefore be neglected in the present discussion if a
non-linear regime is assumed. In contrast, a ``linear'' regime is
also invoked in the vortex creep model for which $\omega_{\rm sp}
<< \omega_{\rm cr}$. However, the above conclusion remains the
same, even for this case, and a "rapid" superfluid spin-up is
again expected until $|\omega| \lesssim \omega_{\rm sp}$ is
achieved! This is because, according to the creep model the
spin-down (up) rate depends "exponentially" on the difference
$(\omega-\omega_{\rm sp})$; see Eq. 28 in Alpar et~al. (1984).
Hence, a relaxation of the superfluid under the assumed conditions
with $\omega_{\rm cr}> |\omega| > \omega_{\rm sp}$ should again
take place on a time scale much shorter than the observed effect
over $\tau_{\rm sp}$, even for the case of a linear regime.

\section{Superfluid Spin-up}
Moreover, the suggested spin-up scenario of APC should be
dismissed at once since the required torque on the superfluid,
during $\tau_{\rm sp}$, may not be realized at all, under the
assumed conditions of (creeping of the) pinned vortices (see
Fig.~1a). That is the pinned superfluid could not be spun {\em up}
by the crust (ie. its container) while the latter is spinning {\em
down}. This is simply because a spinning down vessel (or even one
with a stationary constant rotation rate) albeit rotating faster
than its contained superfluid could not result in any further {\em
spin up} of the vortex lattice which is, by virtue of the assumed
pinning, already {\em co-rotating} with it! As is well-known, an
inward radial motion of the vortices, associated with a spin-up of
the superfluid, requires the presence of a corresponding forward
azimuthal external force acting on the vortices. This is indeed a
trivial fact, considering that any torque on the bulk superfluid
has to be applied primarily on the vortices. However, no {\em
forward} azimuthal force (via scattering processes between the
constituents particles of the vortex-cores and the crust) may be
exerted by the spinning {\em down} crust on the vortex lattice
which is already co-rotating with it. The azimuthal external force
$F_{\rm ext}$, being the viscous drag of the permeating electron
(and phonon) gas co-rotating with the crust, depends on the
relative azimuthal velocity $v_{\rm rel}$ between the {\it crust}
and the {\it vortices}, as well as the associated
velocity-relaxation timescale $\tau_v$ of the vortices. The
external drag force, per unit length, is given as
\markcite{AS8,JM8}
\begin{eqnarray} n_{\rm v} F_{\rm ext} & = &  \rho_{\rm c} {v_{\rm
rel} \over \tau_v },
\end{eqnarray}
where $n_{\rm v}$ is the number density of the vortices per unit
area, and $\rho_{\rm c}$ is the effective density of the
``crust''. Hence, for a spin-up of the superfluid to be achieved,
the crust may impart the corresponding torque on the vortices only
if it (tends to) rotates faster than the vortices, so that $v_{\rm
rel}$ points to the proper forward direction. Accordingly, a
superfluid spin-up requires the crust to be itself spinning up, or
else if the crust is spinning down, the vortices must be already
rotating slower than the crust; a requirement which is against the
pinning condition. It should be trivially clear that the Magnus
force could not be responsible for the required external torque,
as it is an internal force exerted by the superfluid itself.
Therefore, {\em no further superfluid spin-up} might be expected
to occur during the interval $\tau_{\rm sp}$, namely after the
initial fast spinning up of the crust, as well as the superfluid
and its {\em vortices}, has been accomplished during the glitch
rise time (compare Fig.~1a with Fig.~1b).

The suggested long-term (over time $\tau_{\rm sp}$) superfluid
spin-up in APC is a generalization of the vortex creep model to
the case of a {\em negative} lag, in contrast to the usual
applications of the model to spin-downs driven by a positive lag.
However, according to the existing formulation of the vortex creep
model, a spinning up of the superfluid would require a {\it
positive} accelerating torque $N_{\rm em}$ acting on the whole
star (see, eg., Eqs.~28, and 38 in Alpar et~al. 1984). Application
of the same formalism (as is attempted in Eq.~5 of APC) to the
suggested case of an spin-up in presence of the given {\it
negative} $N_{\rm em}$  is not, a priori, justified; it is indeed
contradictory.

The vortex creep model suggests that a radial Magnus force, due to
a superfluid rotational lag, results in a radial bias in the
otherwise randomly directed creeping of the vortices
\markcite{AAP} (Alpar et~al. 1984; see Jahan-Miri 2005 for a
critical discussion of the vortex creep model on this, and other,
grounds). This might be, mistakenly, interpreted to imply that
given a negative lag the inward creeping motion of the vortices,
hence a superfluid spin-up, should necessarily follow,
irrespective of the presence or absence of the needed torque on
the superfluid. As already noted, the role of driving the vortices
inward, ie. spinning up of the superfluid, may not be assigned to
the Magnus force. The Magnus force associated with the rotational
lag is a {\em radial} force and is also an internal force exerted
by the superfluid {\em itself}; both properties disqualifying it
from being the source of a torque on the superfluid. Accordingly,
the point to be emphasized is that the obvious requirement for a
spin-up process, namely the realization of the needed torque, is
indeed missing in the suggested mechanism of APC. Moreover, the
inability of the superfluid to be spun up, in this case, is {\it
not} a direct consequence of the fact that pinned vortices may not
respond freely to an applied external torque. Rather, the vortices
under the assumed conditions do not have any ``tendency'' for an
inward radial motion, in spite of the presence of an inward radial
Magnus force which is balanced by the pinning forces. Thus
creeping motion of the vortices may not be invoked as a
resolution; radial creep should be prohibited accordingly.  A
change in the spin frequency of a superfluid involves not only a
radial motion of the vortices but also a corresponding azimuthal
one, as may be also verified from the solution of equation of
motion of vortices during a superfluid rotational relaxation (see
Eq.~9 in Alpar \& Sauls 1988, and Eq.~4 in Jahan-Miri 1998). The
torque may be transmitted only during such an azimuthal motion and
would nevertheless require and initiate a radial motion as well.
Therefore, purely {\it radial} creeping of the vortices, which is
implied by the existing formulation of the vortex creep model, may
not be invoked as a spinning-up mechanism during the transition
from \(\Omega_{\rm sp} - \Omega_{\rm L}= -\omega_{\rm sp}\) to
\(\Omega_{\rm sp} \gtrsim \Omega_{\rm L}\). That is to say, no
radial (creeping) motion of the vortices is permitted, for the
assumed case, because the spinning down crust could not impart any
{\em forward} azimuthal force on the pinned vortices. The
superfluid would rather remain decoupled at a constant value of
$\Omega_{\rm sp}$ (if not spinning down) during this transition
which is achieved due only to the spinning down of the crust, as
depicted in Fig.~1b.

\section{Concluding}
Decoupling of (a part of) the moment of inertia of the crust of a
neutron star at a glitch, from the rest of the co-rotating star,
could readily account for the excess post-glitch spin-down rates
comparable to the fractional moment of inertia of the decoupled
part. The same preliminary fact applies to a (partial) decoupling
of a (pinned) superfluid component in the crust, or elsewhere in
the star, as well. A decoupled component (say, in the crust)
could, in principle, result in an even larger excess spin-down
rate of the star if it is further assumed to be spinning up while
the rest of the star is spinning down; ie. a {\em negative}
coupling instead of a mere decoupling. Nevertheless, here we have
shown that the only suggested mechanism for such a {\em negative}
coupling of a pinned superfluid part in the crust not only fails
quantitatively to account for the observed effect in pulsars, it
is also ruled out conceptually since the required torque on the
superfluid could not be realized at all.

Given the standard picture of the interior of a neutron
star\markcite{S89,P92}(Sauls 1989; Pines \& Alpar 1992), one is
thus left to speculate on the possible role of the stellar core to
induce the observed effect. That is, the observed large spin-down
rates, over timescales of a day and more, should be caused by a
decoupling of (a part of) the stellar core. The large moment of
inertia $I_{\rm core}$ of the core, with \( {I_{\rm core} \over I}
\sim 90~\% \), may easily account for the observations.
Nevertheless, a non-superfluid component in the core would couple
to the crust on very short timescales ($< 10^{-11}$~s)
\markcite{BP9}(Baym et~al. 1969a), and could not have any
footprint left in the observed post-glitch relaxation. Also, a
core-superfluid with free (unpinned) vortices would be again
expected to have very short coupling timescales of the order of
less than one or two minutes \markcite{AS8,P92,JM8}(Alpar \& Sauls
1988; Pines \& Alpar 1992; Jahan-Miri 1998). However, the pinning
of the superfluid vortices to the superconductor fluxoids in the
core of neutron stars, might offer a way out of the dilemma. The
effect has been originally suggested on theoretical grounds
\markcite{MT,S89,JO91b}(Muslimov \&  Tsygan 1985; Sauls 1989;
Jones 1991b), while its observable consequences for the rotational
dynamics of a neutron star as well as its magnetic evolution have
been investigated by various authors
\markcite{SB90,JO91b,CCD,JM6,RZC,JM00,JM02}(Srinivasan 1990; Jones
1991b; Chau, Cheng \& Ding 1992; Jahan-Miri 1996; Ruderman, Zhu \&
Chen 1998; Jahan-Miri 2000; 2002).

Very briefly, the peculiar feature of the assumed pinning in the
core, that is the {\em moving} nature of the {\em pinning sites},
has to be highlighted, in this regard. The fluxoids are indeed
predicted to undergo a steady outward radial motion throughout the
active lifetime of a pulsar \markcite{CCD,JM0} (Chau, Cheng \&
Ding 1992; Jahan-Miri 2000). At a glitch, a departure in the lag
from its earlier critical value, causes a (dynamically partial)
decoupling of the core, which may explain the observed initial
large spin-down rates soon after a glitch (that could not be
possibly caused by the crust, as discussed above). Nevertheless
the core superfluid does spin down even before the critical lag
value is restored, simply because the pinning barriers (the
fluxoids), hence the superfluid vortices pinned to them, are
moving radially, at all times. Otherwise, for the core superfluid
to remain completely decoupled (the vortices being stationary) the
superfluid rotation lag should amount to its critical value! since
a crossing-through of the vortices and fluxoids, ie. unpinning
events, would be inevitable. For the core superfluid, a {\em
decoupling} accompanies and implies {\em unpinning!}, contrary to
the usual case of stationary pinning sites, as for the crust.
Hence, during a post-glitch recovery phase (while the lag is still
far from its critical value) the vortices move out along with the
fluxoids, keeping the superfluid in a spinning down state, albeit
at a slower rate than before the glitch. A quantitative modelling
of the effects of a pinned superfluid component in the core on the
post-glitch response, as well as its role in a (free) precession
of the star, remains to be further studied in details.

This work was supported by a grant from the Research Committee of
Shiraz University.

\begin{figure}
\epsscale{2.4}
\vspace{-5cm}
%\hspace{-10cm}
\plotone{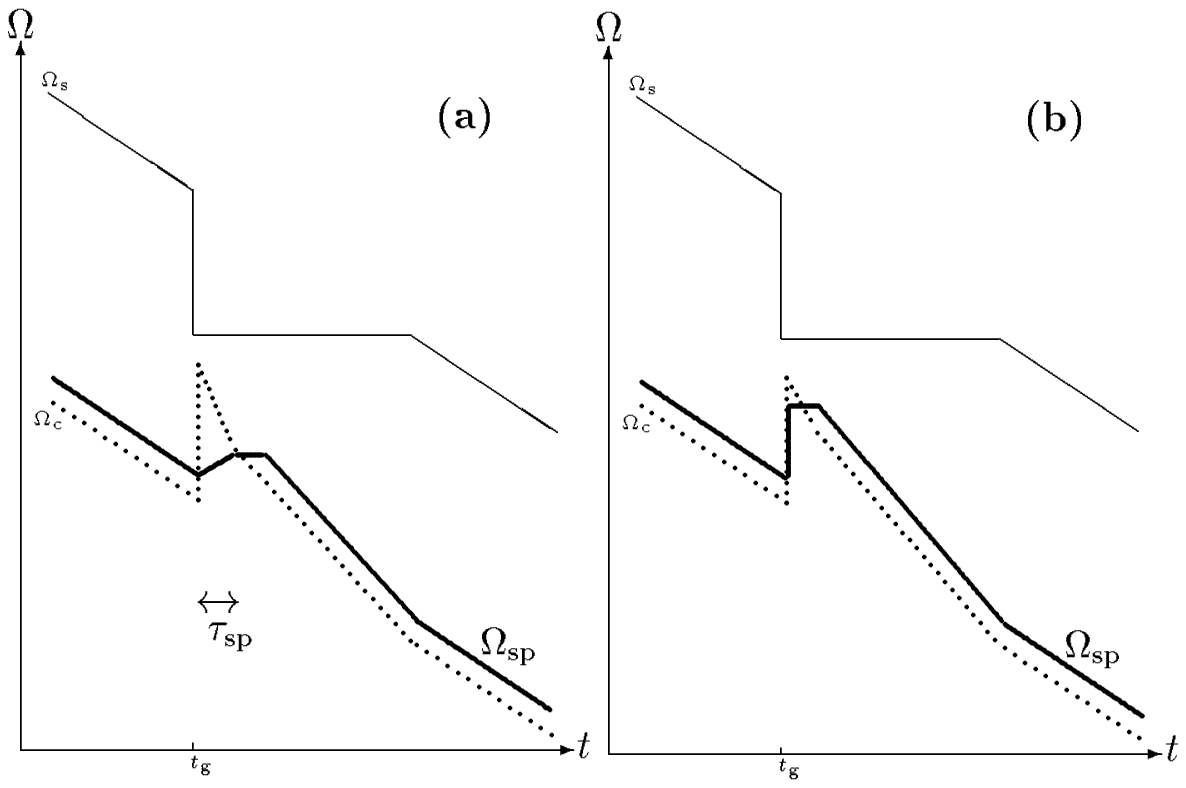}
\vspace{-15cm}
\caption{As is
shown, schematically, by the {\it thin} and the {\it dotted}
lines, a superfluid, as in the crust of a neutron star, whose
vortices are pinned and co-rotating, at a rate $\Omega_{\rm c}$,
with its spinning down vessel, may spin-down at the same rate as
the vessel provided a certain lag between its rotation frequency
$\Omega_{\rm s}$ and $\Omega_{\rm c}$ is maintained. A decrease in
the lag, say at a glitch, results in a decoupling of the
superfluid and hence a larger spin-down rate for the vessel. The
superfluid starts spinning down again once the lag is increased to
its steady-state value. The {\it thick} line represents the
superimposed behavior of a second superfluid component,
$\Omega_{\rm sp}$, which supports a much smaller steady-state lag.
Its behavior as implied in APC90 is shown in the panel {\bf (a)},
in contrast to that suggested here, shown in {\bf (b)}. The
unresolved rising time of $\Omega_{\rm c}$ at the glitch instant
$t=t_{\rm g}$ should be reckoned as a measure of the expected
coupling time of the superfluid due to free motion of vortices, in
contrast to the suggested spin-up period $\tau_{\rm sp}$ in {\bf
(a)}.} \label{f1}
\end{figure}

\end{document}